\pdfoutput=1
\documentclass[aip,jap,reprint,superscriptaddress,amsmath,amssymb]{revtex4-1}
\usepackage{epsfig,psfrag}
\usepackage{textgreek}
\usepackage{dcolumn}
\usepackage{bm}
\usepackage{graphicx}
\usepackage{color}
\usepackage{epstopdf}
\usepackage{textcomp}
\graphicspath{{./Images/}}
\newcommand{\MF}[1]{#1}

\begin{document}
	\title{Thermoelectrically cooled THz quantum cascade laser operating up to 210 K}
	
	\author{L. Bosco}\email{lbosco@phys.ethz.ch}
	\author{M. Francki\'e}
	\author{G. Scalari}
	\author{M. Beck}
	\affiliation{Institute for Quantum Electronics, ETH Zurich, Auguste-Piccard-Hof 1, 8093 Zurich, Switzerland}
	\author{A. Wacker}
	\affiliation{Mathematical Physics and Nanolund, Lund Unviversity, Box 118, 221 00 Lund, Sweden}
	\author{J. Faist}
	\affiliation{Institute for Quantum Electronics, ETH Zurich, Auguste-Piccard-Hof 1, 8093 Zurich, Switzerland}
	
	\begin{abstract}
		We present a \MF{terahertz} quantum cascade laser operating on a thermoelectric cooler up to a record-high temperature of 210.5 K. The active region design is based on only two quantum wells and achieves high temperature operation thanks to a systematic optimization by means of a  nonequilibrium Green's function model. Laser spectra were measured with a room temperature detector, making the whole setup  cryogenic free. At low temperatures ($\sim 40$ K), a maximum output power of 200 mW was measured. 
	\end{abstract}
	\maketitle
\begin{figure}[!ht]
		\includegraphics[scale=0.5]{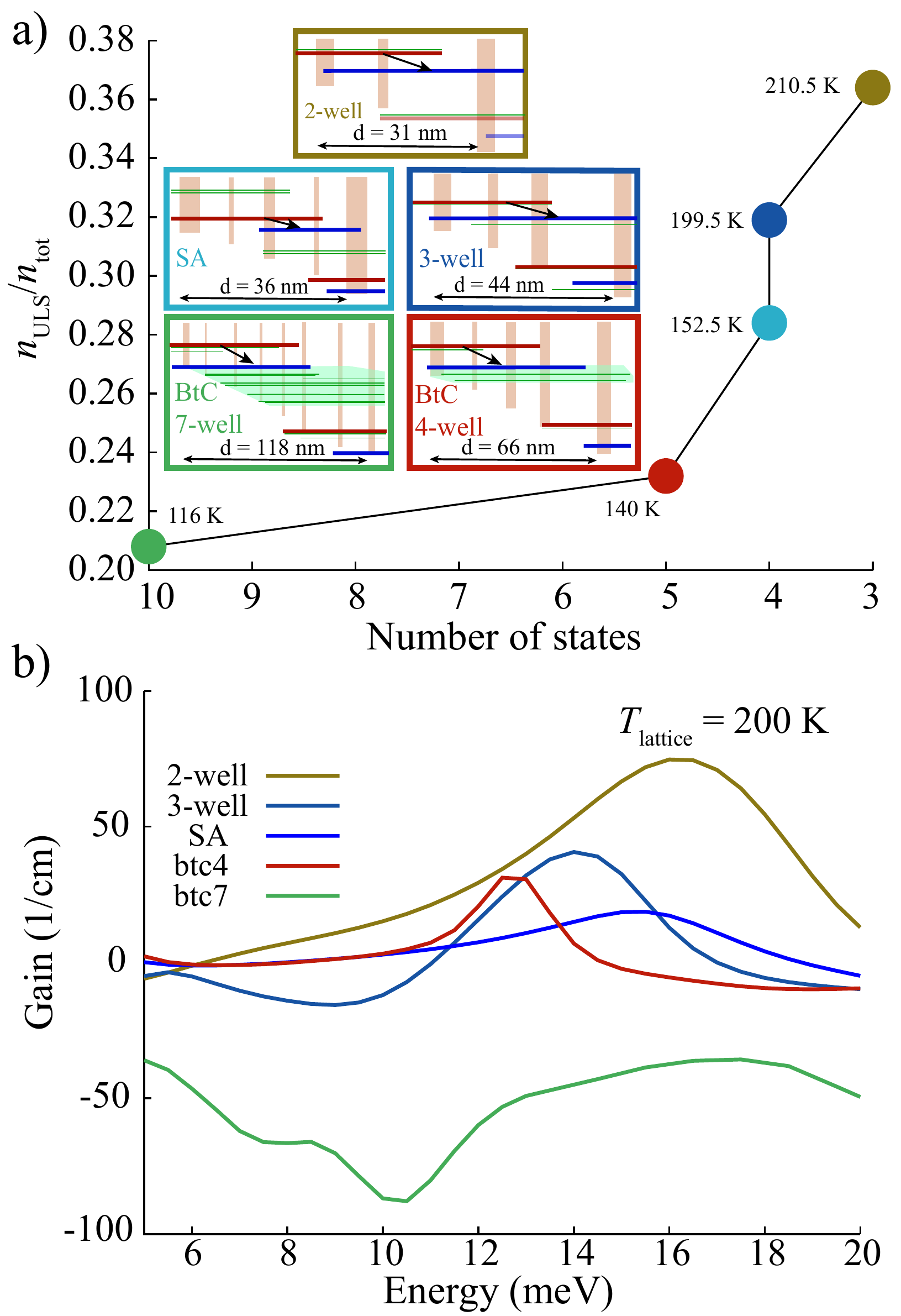}
		\caption{(a) Relative density of the upper laser state ($n_\text{ULS}$) calculated with a nonequilibrium Green's function (NEGF) model for five THz quantum cascade laser designs with different number of active states per period. Schematics of the simulated QCLs are shown in the five insets and represent 2-well, 3-well\cite{FathololoumiOE2012}, scattering-assisted (SA) injection\cite{RazavipourJAP2013}, 4-well\cite{AmantiNJP2009} and 7-well\cite{ScalariAPL2005} bound-to-continuum (BtC) designs. The 2-well QCL is the layer presented \MF{in this work}. For each design we show the upper laser state (ULS) in red, the lower laser state (LLS) in blue, and green lines indicate the other relevant states. The vertical bars show the barriers separating the quantum wells. (b) Comparison of simulated gain for the five designs. The fewer the states, the more concentrated the carriers are to the ULS, making gain exceed re-absorption.}
		\label{fig:1}
	\end{figure}

	\MF{Terahertz (THz)} radiation is subject to a wide range of research and technological efforts\cite{DhillonPRD2017}, from solid state fundamental physics to biomedicine and astrophysics. Specifically, since  materials such as textiles, plastics, coatings and biological tissues are transparent to THz radiation, this spectral region is also promising for a variety of non-invasive imaging and non-destructive quality assessment applications such as airport security screening and thickness coating measurements.
	In order to unlock the full potential of these applications on a large scale, compact and powerful THz sources are needed. A promising candidate is the quantum cascade laser (QCL)\cite{FaistScience1994,KohlerNature2002}, a compact injection laser based on semiconductor heterostructures. THz QCLs have already shown  high emitted powers (several hundred mW both in pulsed and continuous wave) \cite{WilliamsNatPhotonics2007, LiEL2014, BrandstetterAPL2013} and spectral coverage throughout the $1$-$6$ THz range\cite{SirtoriNatPhot2013} with single mode and broadband devices\cite{RoschNatPhot2014}. However, despite several efforts, to-date THz QCL operation is still restricted to cryogenic cooling \MF{and} their maximum operating temperature is still below 200 K \cite{FathololoumiOE2012}. 
	In this Letter we present a THz QCL operating \MF{up to 210 K}, allowing the use of a small footprint, 4-stage Peltier cooler. In combination with a commercial DTGS detector, this constitutes a platform for THz spectroscopy completely free from any cryogenic and \MF{Helium-based (He-based)} technology. 
	For any given wavelength, resonator losses, and broadening lifetime\MF{,} the ultimate temperature limit to QCL laser operation increases with the fraction of the electron population in the upper state \cite{FaistAPL2007}, as parasitic reabsorption by the free electrons participating to the transport and maintaining the electrical stability are the key source of optical losses\cite{FaistEL2010}. This line of reasoning provides an explanation for the general trend of THz QCL designs that has been towards reducing the number of states per period. This qualitative result is supported by the quantitative gain computations shown in Fig.~\ref{fig:1}(a), where the computed active region population and gain at $200$ K of a selection of THz QCLs using different design schemes show a relative increase in the upper laser state (ULS) population ($n_\text{ULS}$) as the number of active states is reduced. This leads to both an increase in population inversion and a reduction of intersubband re-absorption and justifies using this quantity as a performance indicator\cite{FaistAPL2007}. Early THz QCLs used the "bound-to-continuum" (BtC) scheme\MF{\cite{KohlerNature2002, ScalariAPL2003, ScalariAPL2005, AmantiNJP2009}}, where the upper laser state (ULS) is localized in one quantum well and population inversion is achieved by a combination of intersubband scattering and tunneling transitions. The main limitation resides in the re-absportion of photons by thermally populated states in the lower miniband, as evident by the negative gain in Fig.~\ref{fig:1} (b). To avoid this problem,  the "resonant phonon" 3-well scheme achieves efficient LLS extraction using only four states\MF{\cite{LuoAPL2007, KumarAPL2009}}, by exploiting a single tunneling transition followed by resonant phonon emission. The scattering assisted (SA) injection design\cite{DupontJAP2012}, also with 4 active states per period, achieves similar performance. The minimum number of active states is three, limited by the electrical stability of the laser\cite{FaistEL2010, WackerAPL2010}, and has been realized in two-quantum well designs\MF{\cite{ScalariITQW2009,ScalariOE2010,KumarAPL2009B}} with close-to-record temperature performance achieved recently\cite{FranckieAPL2018, AlboAPL2017}. In fact, one might see the two quantum well active region as the structure closest to the original proposal from Kazarinov and Suris\cite{KazarinovSPS1971} which satisfies the electrical stability condition. The analysis in Fig.~\ref{fig:1} \MF{compares the best devices obtained from different designs regardless to the emitted frequency and} indicates that using the minimum number of states per period should lead the choice of design scheme for improving THz QCL temperature performance.
	
\begin{figure*}[!ht]
		\includegraphics[scale=0.70]{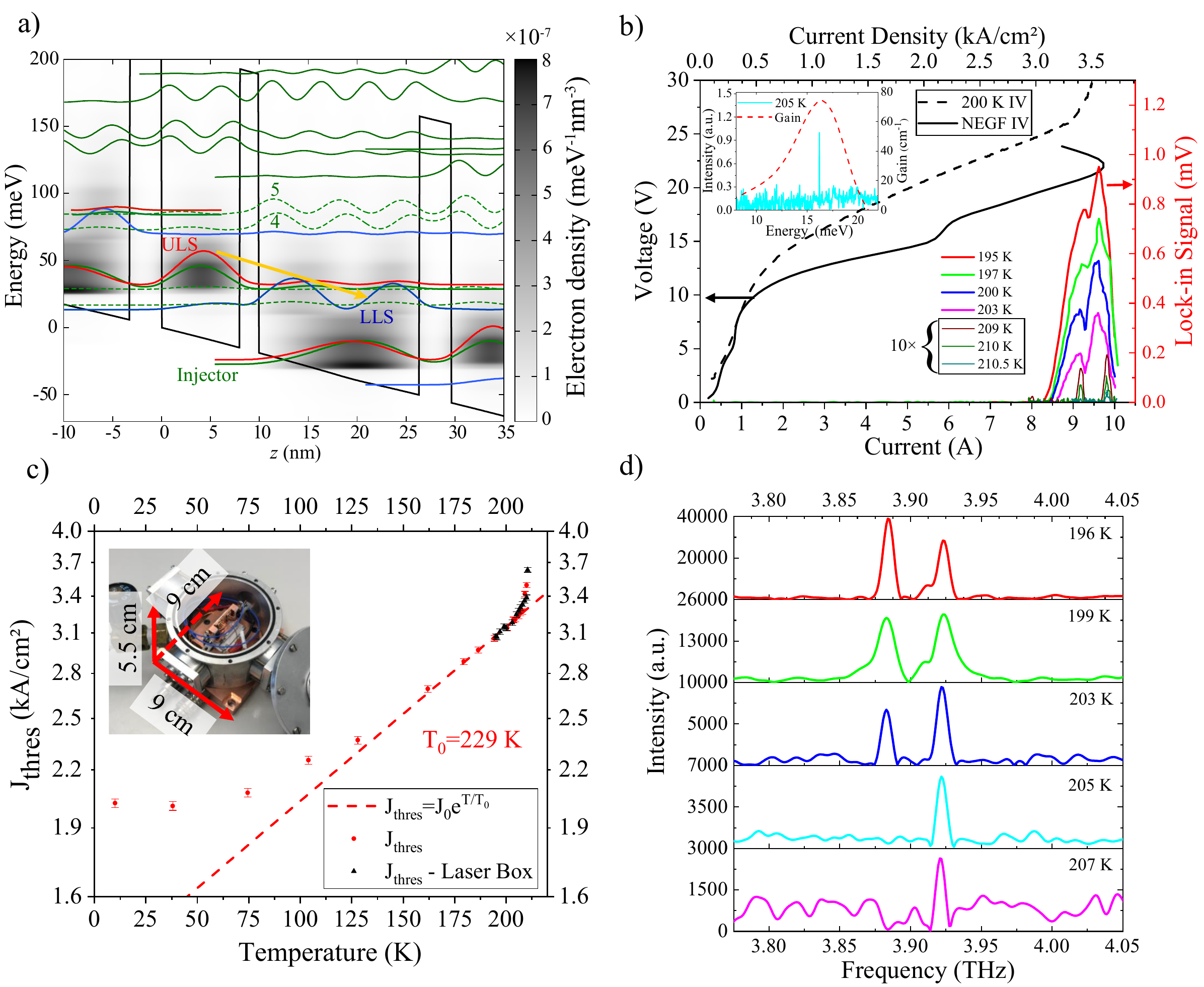}
		\caption{a) Band structure of the GaAs/Al${}_{0.25}$Ga${}_{0.75}$As QCL with highest operating temperature with layer sequence (in \AA~ with barriers in bold face) \textbf{32.6}/79.9/\textbf{19.0}/84.0/\underline{29.0}/51.6 where the underlined well section is doped to $n=4.5\times10^{10}$ cm${}^{-3}$, shown at a bias of 56 mV/period where the current density is maximal. The yellow arrow indicates the diagonal ($f_\text{osc.}=0.20$) laser transition from the upper (red) to the lower (blue) laser state. The green dashed lines indicate states in close resonance to the lower and upper laser states, respectively. The grey scale shows the electron density evaluated in the NEGF model at a lattice temperature of 200 K. b) LI curves obtained cooling down the laser in the laser box and detecting it using a He-cooled Ge bolometer. In dashed and solid black lines, a comparison between the measured IV and the simulated IV at a lattice temperature of $200$ K is shown. The IV has been measured in a flow cryostat setup. In the inset, it is possible to see that the emitted frequency at $205$ K fits the simulation prediction of the gain. c) $T_0$ fit of the threshold current density $J_{\mathbf{thres}}$ reported on a log scale. Fit is executed on data points collected in cryostat operation (red circles), but agrees well with the data collected during the thermoelectric cooling operation. A $T_0=229$ K is obtained. In inset, the laser box with Peltier cell and mounted QCL. The box footprint ($\sim 9\times9\times5.5$ cm$^3$) is significantly reduced compared to a flow cryostat (even without taking into account the He dewar). At maximum capacity, the Peltier cell uses $\lesssim30$ W to cool the laser down to $195$ K. d) Spectra taken with the laser cooled down using a thermoelectric multistage cooler and using the DTGS room temperature detector of a Bruker Vertex $80$ FTIR. Laser has been operated at maximum power.}
		\label{fig:image_3}
	\end{figure*}

	Nonequilibrium Green's function (NEGF) modeling has proven to be a powerful tool for optimizing THz QCL designs\cite{FranckieAPL2018}. \MF{Owing to the significant computation times that come with such a complex model, we have performed a systematic optimization using the NEGF model of Ref.~\onlinecite{WackerIEEEJSelTopQuant2013} in conjunction with an information algorithm with parallel trials for multi-variable, global extremum search\cite{SergeevUSSRCMMP1989} in order to increase the convergence rate. We optimized the maximum gain at a lattice temperature of 300 K}, varying all four layers of the two-well design in a region around the best one from Ref.~\onlinecite{FranckieAPL2018}. In order to keep a manageable computing time, we limited the optimization to a photon energy of $\hbar\omega=16$ meV and a bias of $52.5$ mV/period (encompassing one LO phonon and one photon emission per period).
	We find that the gain is maximized for a shorter period length and thicker barriers compared to the nominal structure previously implemented.
	In particular, a reduction of $\sim3\%$ in the period length is predicted to increase the gain from $\sim18$ to $\sim25$ cm$^{-1}$.
	\begin{figure}[!ht]
		\includegraphics[scale=0.31]{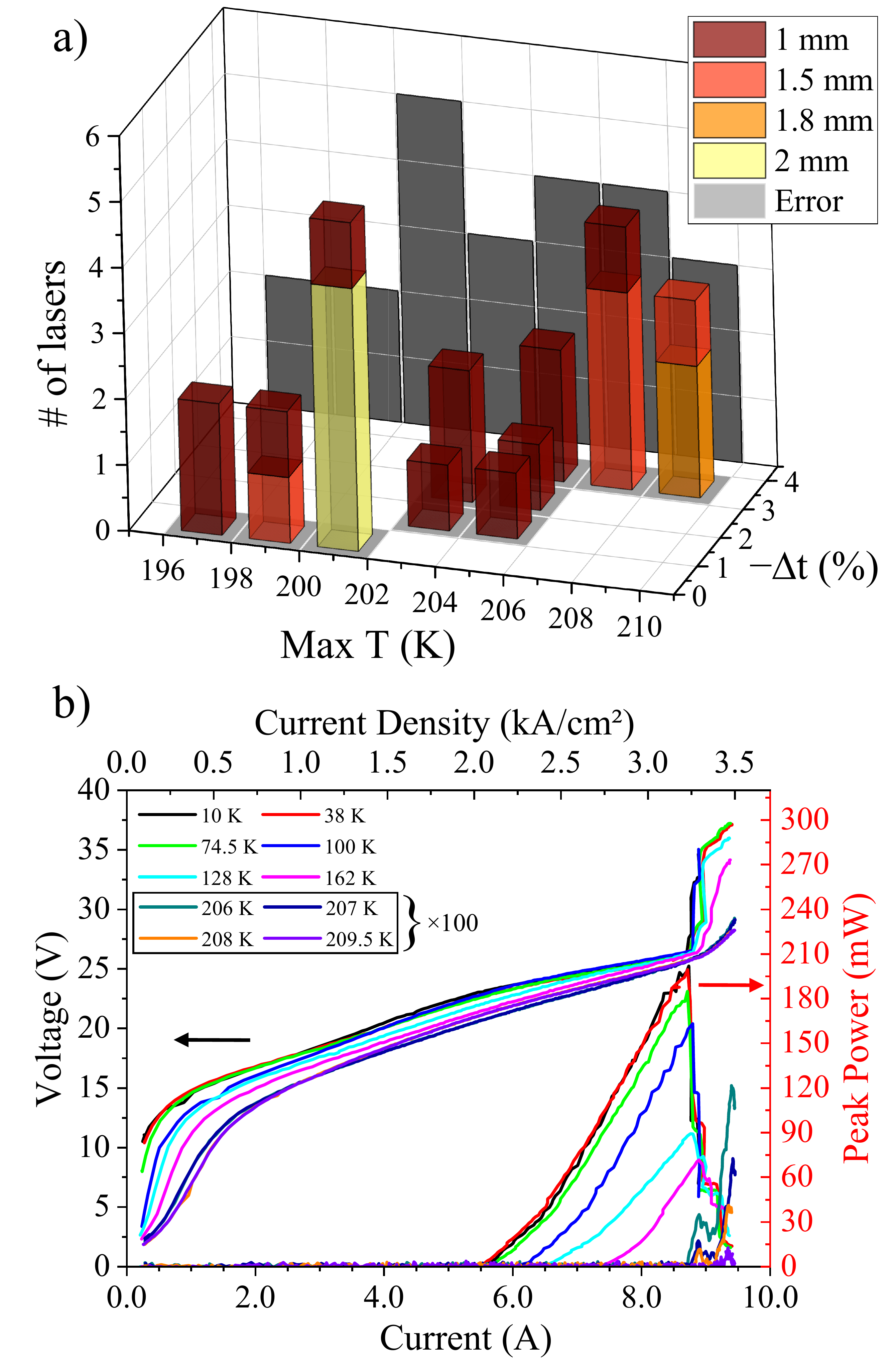}
		\caption{
			a) $3$D histogram of the statistics conducted over $23$ lasers with a width of $150$ $\mu$m and a length between $1$ and $2$ mm, as specified by the color scale, versus the layer $\Delta t$. The projection in dark grey on the XY plane sums over all the thickness variation. b) LIV characteristic of the best device is shown, pumped electrically with $50$ ns width pulses at $415$ Hz, up to $\sim 210$ K.}
		\label{fig:image_1}
	\end{figure}
	The best design as predicted by the NEGF optimization has been grown by molecular beam epitaxy (MBE). 
	As compared to our previous work\cite{FranckieAPL2018}, a thicker ($12$ $\mu$m) active region was grown in order to reduce the waveguide losses at the cost of slightly increased mirror losses for a given cavity length\cite{KohenJAP2005}.	
	The grown QCLs were processed as Cu-Cu waveguides \MF{using} a thermocompressive bonding after the evaporation of $5$ nm of Ta and $500$ nm of Cu on the active region.
	As a top metal cladding, a layer of Ta\slash Cu\slash Ti\slash Au ($5/50/20/200$ nm) has been deposited and the sample has been dry etched in a Cl$_2$\slash Ar plasma using SiN$_x$ as hard mask.
	Since the epitaxial layers grown on the  $3"$ wafer exhibits a thickness gradient of approximately $\Delta t = -4\%$ from the center to the edge \MF{(calibrated through X-ray mapping)}, a period thickness span of $\sim0-4\%$ compared to the nominal design is experimentally realized. \MF{The active region thickness was measured by high-resolution x-ray diffraction performing a 5 mm mesh mapping over the 3" wafer. }A systematic characterization of the processed lasers in a He-cooled flow cryostat has confirmed the strong dependence of the laser performance on the active region thickness predicted by the simulations.
	The $3$D histogram in Fig.\ref{fig:image_1}(a) reports the maximum temperature of all the $23$ tested lasers with a width of $150$ $\mu$m and lengths between $1$ and $2$ mm with varying period thickness, exploring ranges of $\pm0.5\%$. As confirmed by the simulations, the maximum operating temperature ($T_\text{max}$) improves when moving towards the edge of the wafer. The best device, found for $\Delta t \approx -3.5\%$, was $1.8$ mm long and lased up to a maximum temperature of $\sim210$ K, as shown in Fig.\ref{fig:image_1}(b). \MF{The structure sequence for this device is \textbf{32.6}/79.9/\textbf{19.0}/84.0/\underline{29.0}/51.6, here expressed in \AA, with barriers in bold face and the underlined well section doped to $n=4.5\times10^{10}$ cm${}^{-2}$. The} band structure and energetically resolved carrier density of this device simulated with the NEGF model are shown in Fig.~\ref{fig:image_3}(a).
	\MF{The main difference with regards to the nominal design\cite{FranckieAPL2018} is a higher extraction energy $E_\text{ex} \approx 41$ meV and a slightly lower oscillator strength (although the latter is highly bias-dependent and therefore difficult to precisely define). This reduces non-radiative phonon emission from the upper laser state as well as thermal backfilling, improving the temperature resilience.}
	
	The laser was electrically driven with $50$ ns wide pulses at a repetition frequency of $415$ Hz. A pre-calibrated He-cooled Si bolometer has been used as a detector showing a peak power of $\sim 200$ mW and a threshold current density of $2$ kA/cm$^2$ when the laser is operated at $10$ K. \MF{The relatively high current density required for laser operation is a result of the optimization of the presented QCL being restricted to high-temperature operation and high output power. In order to limit the current density, the parasitic leakage into level 5 (discussed in some detail in Ref.~\onlinecite{FranckieAPL2018}) could be reduced. However, for the purpose of pulsed high-temperature operation, we find that this current channel has a small impact since less than 10\% of the carriers reside in levels 4 and 5 even at 200 K.}
	
	As shown in Fig.~\ref{fig:image_3}(b-d), the device whose performances are reported in Fig.\ref{fig:image_1}(b)  is then tested on a four-stage cooler 
	(model SP$2394$ from II-VI Marlow Industries), that reached a temperature as low as $195$ K  with a power consumption below $30$ W.  The temperature was measured using a NTC Thermistor ($44000$RC series from TE connectivity). As shown in Fig.\ref{fig:image_3}(b), the light collected using an elliptical mirror is measured at different temperatures from $195$ K up to $210.5$ K as a function of current using a He-cooled Ge bolometer. The results confirm the performance recorded in the flow cryostat, with the difference of $1$ K, corresponding to the accuracy of the silicon temperature sensor in the flow cryostat. The threshold current density exhibits an exponential temperature dependence with a characteristic temperature of $T_0=229$ K. \MF{Close to $J_\mathrm{max}$ the derivative of gain with respect to current density decreases and eventually crosses zero. Thus, in order to reach the threshold gain at temperatures approaching $T_\mathrm{max}$ a rapid increase in $J_\mathrm{th}$ with temperature is observed. The} high value of $T_0$ demonstrates the potential of this active region design for high temperature operation, as a further reduction of $10\%$ of the losses would increase the maximum operating temperature by potentially $20$ K.
	
	As we previously observed in Ref.~\onlinecite{FranckieAPL2018}, the simulated current agrees very well with the experimental one, and the bias is in reasonable agreement. The inset in Fig.~\ref{fig:image_3}(b) shows that also the agreement between the simulated peak gain frequency and the observed emission frequency of the laser are in excellent agreement. This confirms that the NEGF is a powerful tool for optimization of THz QCLs, and the operation of the laser has been \MF{well} understood. The discrepancy between the optimized (nominal) structure and the best ($3.5\%$ thinner) structure (which is also better in the NEGF model under a detailed analysis) can be explained by the restrictions imposed to the optimization, since the best structure operates at a higher bias ($56$ mV/period) and photon energy ($16.5$ meV). In accordance with previous results\cite{FranckieAPL2018}, including a rudimentary form of electron-electron scattering\cite{WingeJPhysConfSer2016} in the model (not shown) the current density changes negligibly while the gain is almost halved and red shifted by 1 meV at 200 K.\MF{We also find that the temperature degradation of the presented design  is dominated by level broadening.\cite{FranckieAPL2018}}

	
	Laser spectra with the device operating on a Peltier cooler were performed using a commercial FTIR (Bruker Vertex 80V) equipped with a DTGS room temperature detector, making such setup fully cryogenic free. Spectra measured at different temperatures with a resolution of $0.2$ cm$^{-1}$ are reported in Fig.\ref{fig:image_3}(a).
	
	In conclusion, we have demonstrated the first operation of a THz QCL using thermo-electric cooling, and the first operation of a THz QCL above 200 K, with a peak power as high as $1.2$ mW at $206$ K, which allowed detection in a cryogenic-free system. The methodology employed in this work will likely lead to further improvements of this technology; pursuing better QCL designs, studying in depth the effect of other parameters, \emph{e.g.} barrier height, doping position, and doping region width, as well as improving the Cu-Cu process is expected to lead to increased maximum operating temperatures and extracted optical power. 
	The presented results pave the way towards a new generation of on-chip, portable THz devices based on high power THz coherent sources. 
	
	\MF{The authors would like to thank Dr. Keita Ohtani and Dr. Christopher Bonzon for their help in the initial stages of this project.} This work is supported by the Horizon 2020 ERC grant project CHIC 724344, the Swedish Research Council (grant 2017-04287), as well as a grant from SFB 956 of the Deutsche Forschungsgemeinschaft. 
	

\end{document}